\documentstyle[preprint,aps,psfig]{revtex}
\begin{document}

\title
{Dynamical properties and plasmon dispersion of a weakly degenerate 
correlated one-component plasma }

\author{ V.~Golubnychiy$^1$, M.~Bonitz$^1$, D.~Kremp$^1$, and M.~Schlanges$^2$
}
\address
{
$^1$ Fachbereich Physik, Universit\"at Rostock,
Universit\"atsplatz 3, 18051 Rostock, FRG
\\$^2$  Institut f\"ur Physik, Ernst--Moritz-Arndt Universit\"at Greifswald, 
Domstr. 10a, 17489 Greifswald, FRG
}
\maketitle

\begin{abstract}
Classical Molecular Dynamics (MD) simulations for a one-component
plasma (OCP) are presented. Quantum effects are included in the form of the Kelbg potential.
Results for the dynamical structure factor  are compared with the Vlasov and
RPA (random phase approximation) theories. The influence of the coupling parameter $\Gamma$, 
degeneracy parameter $\rho { \Lambda }^3$
and the form of the pair interaction  on the optical plasmon dispersion is investigated.
An improved analytical approximation for the dispersion of Langmuir waves is presented.
\end{abstract}


\section{Introduction}\label{ch1}

The model of a classical one-component plasma has - due to its simplicity - been
widely investigated both theoretically and with various
numerical and simulation methods, see e.g. \protect\cite{han73,clwh82,ose97}
and \protect\cite{mat96,shk96}, respectively. Since the pioneering numerical
work of Brush, Sahlin, Teller \protect\cite{bst66}, the thermodynamic and
dynamic characteristics of the classical OCP have been studied in detail.
In particular, the dependence of the properties on the coupling parameter 
$\Gamma = 4\pi e^2 / (\bar{r}k_B T)$,
where $ \bar{r} = ( \frac{3}{4\pi \rho})^{1/3}$ is the mean interparticle distance and 
$\rho$ the density, have
been investigated up to very large values of $\Gamma$ \protect\cite{sdd82,oi87}.
 Among the most important
thermodynamic results is the prediction of crystallization at values of $\Gamma$ of the
order of 172-180 \protect\cite{fh93,i94}. Furthermore, investigations of the dynamic 
properties of strongly correlated classical plasmas have indicated that the wave number 
dependent plasmon dispersion changes from monotonic growth, common for
weakly coupled plasmas, to a decreasing dispersion 
around $\Gamma \approx $ 3 \cite{han73}.

On the other hand, there is growing interest in the dynamic properties of dense {\em quantum} 
plasmas, in particular in astrophysics, laser plasmas and condensed matter. While 
the case of strong 
degeneracy (strong quantum limit) and weak coupling at very high densities is well 
described by the random phase approximation (RPA, see e.g. \cite{pn66,mahan90,kker86,bonitz-book}), 
the properties at {\em intermediate 
coupling and degeneracy} remain poorly explored. Especially, one is interested in the 
dynamic plasma behavior in cases where the average kinetic energy is of the same order 
as the mean potential energy, i.e. $\Gamma \sim 1$, where collisionless theories such 
as the RPA fail, e.g. \cite{mahan90,kker86,kwong-etal.00prl}. For these situations, 
quantum molecular dynamics (QMD) simulations \cite{qmd} are the 
appropriate numerical approach which, however, is yet lacking the required efficiency. For 
weakly degenerate plasmas, with $\rho \Lambda^3 \le 1$, 
where $\Lambda$ is the DeBroglie wave length (see below), it is 
expected that one can perform much simpler classical MD simulations using effective quantum pair 
potentials, e.g. \cite{ose97,kelbg}. These potentials can be  derived from the
2-particle Slater sum using Morita's method. 
It is the aim of this paper to explore this MD approach in detail, 
especially for the analysis of the optical (Langmuir) plasmon dispersion.

It is natural to start this analysis with OCP--simulations because they have the advantage of 
the absence of a collapse of oppositely 
charged particles at small distances. On the other hand, the existence of a homogeneous 
background of oppositely charged particles leads to some additional technical
difficulties compared to 2-component systems, due to restricted carrier rearrangement 
causing less effective screening of the Coulomb interaction. 
One major problem of MD simulations of dynamical properties is that the behavior at 
small wave numbers is difficult to investigate. The reason is that large box-sizes 
are required which, for the analysis of high density plasmas, translates into large 
particle numbers. The current increase of available computer power gives one the possibility to investigate 
size-dependent properties like the density-density correlations 
$\langle \rho_{\vec{k}}(0) \rho_{-\vec{k}}(t) \rangle$ for 
smaller k-vectors than before. In this paper, we are able to present  accurate
results for the dynamical properties of the OCP, such as the dynamical structure factor and the 
wave vector dispersion of Langmuir oscillations. Our simulations for intermediate 
values of the coupling parameter, $\Gamma=1 \dots 4$, show an interesting 
dispersion: the frequency increases up to a maximum and, for large wave numbers,
decreases again.
Further, we investigate the role of quantum effects by comparing simulations 
with the Coulomb potential and an effective quantum pair potential (Kelbg potential
\protect\cite{kelbg}) for the region of small and intermediate coupling.
 We found that quantum diffraction effects have noticable influence on the behavior of 
the optical dispersion curves. Increase of the degeneracy leads to a softenig  of the
dispersion $\omega(k)$, especially at  intermediate  wave vectors. 

\section{Dynamical properties of the OCP}\label{2}
\subsection{Statistical approach}
A central quantity to determine the dynamic properties of charged many-particle systems is  
the frequency-dependent dielectric function $\epsilon(\vec{k}, \omega)$ which, for the 
OCP, is given by 
\begin{eqnarray}
\epsilon(\vec{k}, \omega) = 1 - U_C(\vec{k}) \Pi (\vec{k}, \omega).
\label{df}
\end{eqnarray}
Here $U_C(\vec{k})$ is the spatial Fourier transform of the Coulomb potential, 
$ U_C(k) = 4\pi e^2/k^2$, and
$\Pi (\vec{k}, \omega)$ is the longitudinal polarization function. Thus, many-body effects
enter the dielectric function via $\Pi$.
There exist many  approximations for the latter function, the simplest one being 
mean-field theories which neglect short-range correlation effects, i.e. collisions between 
the particles. For the classical OCP, the mean-field result is the Vlasov polarization:
\begin{eqnarray}
\Pi_{\rm Vlasov} (\vec{k},\omega) =  -\frac{1}{m}\int
\frac{d^3 v}{(2 \pi)^3}
\frac{\vec{k}} {\omega - \vec{k}\vec{v} + i\delta}
\frac{\partial F(\vec{v})}{ \partial \vec{v}}.
\label{p_vlasov}
\end{eqnarray}
Here $\delta \rightarrow +0$, indicating the retarded (causal) character of the polarization 
and the dielectric function. Further, $F$ is the distribution function. 
The Vlasov polarization applies only to classical plasmas, where the wave character of the 
particles can be neglected. Quantum effects are important if the interparticle distance or 
the  Debye radius become comparable to 
the DeBroglie wave length $\Lambda=h/\sqrt{2\pi m k_B T}$. Therefore, quantum 
diffraction effects should show 
up in the dielectric properties at large wave numbers. The quantum generalization of the 
Vlasov polarization is the RPA polarization function given by
\begin{eqnarray}
\Pi_{RPA} (\vec{k},\omega) = - \int \frac{d^3 p}{(2 \pi \hbar)^3}
\frac{f(\vec{p})-f(\vec{p}-\hbar\vec{k})} {\hbar\omega + \frac{p^2}{2m} -\frac{(\vec{p}+\hbar\vec{k})^2}
{2m} + i\delta}.
\label{p_rpa}
\end{eqnarray}
In this paper we consider only plasmas in 
equilibrium, so $F$ and $f$ are the Maxwell and Fermi function, respectively. 
One readily confirms that,
 in the limit of long  wavelengths, $\vec{k} \rightarrow 0 $, 
indeed the RPA result (\ref{p_rpa}) goes over to the Vlasov polarization function (\ref{p_vlasov}).
An important quantity which follows from dielectric function (\ref{df}) via the 
fluctuation-dissipation theorem is the dynamical
structure factor $S(\vec{k},\omega)$
\begin{eqnarray}
S(\vec{k},\omega) = -\frac{k_B T}{\pi U_C(k) \omega} \:
{\rm Im} \frac{1}{\epsilon(\vec{k}, \omega)},
\label{s_mf}
\end{eqnarray}
which shows the frequency spectrum of density fluctuations for
a given value of $ \vec{k}$.

As mentioned above, the mean field expressions (\ref{p_vlasov}) and (\ref{p_rpa}) neglect 
short-range correlations and are, therefore, valid only for weakly coupled plasmas, 
$\Gamma \ll 1$. There exist many theoretical concepts to go beyond the RPA which are based 
on quantum kinetic theory, density functional theory and other approaches. This is beyond 
the scope of this paper, see e.g. 
Ref. \cite{mahan90,kker86,kwong-etal.00prl} and references therein. Here, we consider the alternative 
approach to the OCP at finite coupling which is based on molecular dynamics simulations.

\subsection{Molecular dynamics approach to the dynamical properties}

The dielectric and dynamical properties of an interacting many-particle system are 
easily accessible from the density-density correlation function which is defined as
\begin{eqnarray}
A(\vec{k},t) = \frac{1}{N} \langle\rho_{\vec{k}}(t) \rho_{-\vec{k}}(0)\rangle,
\label{dcf}
\end{eqnarray}
where $N$ is the number of particles. $\rho_{\vec{k}}(t)$ is the Fourier component of 
the density,
\begin{eqnarray}
\rho_{\vec{k}}(t) = \sum_{i=1}^{N}  e^{i \vec{k} \vec{r_i}(t)},
\label{rho_k}
\end{eqnarray}
which is computed from the trajectories $\vec{r}_i(t)$ of all particles. 
The dynamical structure factor is just the Fourier transform of the density-density correlation 
function (\ref{dcf})
\begin{eqnarray}
S(\vec{k},\omega)= \frac{1}{2 \pi} \int\limits_{-\infty}^{+\infty} dt \, 
e^{i\omega t} \, A(\vec{k},t). 
\label{s_md}
\end{eqnarray}
Equation (\ref{s_md}) can be directly compared to formula (\ref{s_mf}) and, thus, 
allows for a comparison of the simulation results with the statistical theories. 
Furthermore, Eq. (\ref{s_md}) allows to investigate the influence of quantum effects 
on the dynamical properties and plasmon dispersion of the OCP. Variations of the 
interaction potential (see below) directly affect the particle trajectories and, 
via Eqs. (\ref{dcf})--(\ref{s_md}), the dynamical structure factor. 

\section{Details of the MD-simulations}\label{ch3}

The simulations have been performed in a cube of length $L$ containing $N$ interacting 
electrons on a uniform positive background. For this system, we solved Newton's
equations of motion containing all pair interactions which are derived from a total 
potential $U(r)$, see below.
As an algorithm of motion we used a second-order scheme in form of the Swope algorithm 
\protect\cite{sabw82}.
Since our simulations are performed in the microcanocial ensemble, the mean kinetic 
energy may change. Therefore, to maintain the chosen value of temperature and $\Gamma$, 
we applied scaling (renormalization) of all velocities at every second step.

A central goal of our simulations was to study the influence of quantum effects. We, 
therefore, performed several simulations which used either a Coulomb potential or 
an effective quantum pair potential (see below). To permit flexibility in the use of the 
potential, $U$ was divided into a short-range and a long-range part, 
$U=U^{\rm sr}+U^{\rm lr}$, where quantum effects influence only $U^{\rm sr}$, whereas the 
behavior at large distances, $U^{\rm lr}$, is dominated by the long-range Coulomb interaction.
Let us first describe the treatment of the long-range term.

\subsection{Long-range interaction}
The long-range interaction was computed in standard way using periodic boundary conditions 
and the Ewald summation procedure 
\protect\cite{nd57,sd76}. As a result, the long-range potential is given by the Coulomb 
interaction in the main box and all image cells:
\begin{eqnarray}
 U^{\rm lr}(\vec{r}) &=& 4\pi e^2 \sum^N_{i\ne j}V_{\rm Ewald}(\vec{r}_{ij}),
\label{ulong}
\\ 
V_{\rm Ewald}(\vec{r}) &=& \sum^{n_x,n_y,n_z \le 1}_{\vec{n} =0} 
\frac{{\rm erfc}\left[\sqrt{\pi} |(\vec{r} +\vec{n}L)/L|\right]}{|\vec{r}+ \vec{n} L|}
 + \sum^{n_x,n_y,n_z \le 5, n^2 \le 27}_{\vec{n}\neq 0} \frac{{\rm exp}(-\pi n^2) 
\cos(2 \pi \vec{n} \vec{r} /L)}{\pi n^2 L} - \frac {1}{L},
\label{ewald}
\end{eqnarray}
where erfc is the complementary error function, $L$ -
the side length of the simulation cell and $\vec{n}$ - a vector of integer numbers which 
labels the periodic images of the simulation box.
In this  expression, the first term corresponds to a potential of particles with
Gaussian broadened charge distribution around the electrons with a width of $\sqrt{\pi}$, 
the second one - the compensating Gaussian distributions, and the last one accounts for
the influence of the homogeneous background. It turns out that the second term in (\ref{ulong}) 
can be reduced to 2 loops (one over the particles and one over the vectors $\vec{n}$ 
in the reciprocal 
space) and is not very time consuming. 
The more complicated part is the first term which contains three loops. In case of 
a two-component plasma, a proper choice of the width of the Gaussian distribution and
use of periodic boundary conditions greatly simplifies this term due to cancellations. In 
contrast, for an OCP, the background cancels the interactions only partially, ``statically''. 
As a result, convergence of the sum is  worse, and one needs to take into account all first 
neighboring image cells (total of 26) at every time step. 
 The contribution of all neighboring cells except for the main one  
($0 < |\vec{n}| \le \sqrt{3} $) was computed, before the start of the simulations 
and stored in  3-dimensional tables for the 
potential and forces. During the simulations, we
used 3D-bilinear interpolation at every step to obtain the values of the potential and 
forces for intermediate distances. We found that 100 grid points in every direction are 
adequate, so the total size of the table was $10^6$ elements.
The particle interactions inside the main ($\vec{n} = 0$) cell were evaluated directly at 
every time step without minimum image convention.

\subsection{Short-range interaction. Quantum effects}
Let us now discuss the short-range potential. As has been shown by Kelbg and co-workers 
\cite{kelbg,ebeling}, quantum effects can be treated efficiently by an 
effective pair potential, the Kelbg potential:
\begin{eqnarray}
U_{\rm KELBG}(r,T)=4\pi e^2 \!\left(\frac{1-\exp(-r^2/\lambda^2)}
{r}+\frac{\sqrt\pi}{\lambda}{\rm erfc}(r/\lambda)\right)
\label{kelbg}
\end{eqnarray}
where $\lambda=\frac{\Lambda}{\sqrt{2\pi}}$. As a consequence of quantum effects, this 
potential differs from the Coulomb 
potential at small distances $r\le \lambda$ and is finite at $r=0$. Further, it is 
temperature-dependent via the thermal DeBroglie wavelength.
The Kelbg potential can be regarded as the proper quantum pair potential following from
the two-particle Slater sum $S_2$ without exchange effects:
\begin{eqnarray}
ln S_2 = - \frac{U_{KELBG}}{kT} + O({\Gamma}^2).
\label{slat_sum}
\end{eqnarray}
It treats quantum diffraction effects exactly, up first order in $\Gamma$.
 Frequently other quantum pair potentials  have been used, including the Deutsch 
 potential \protect\cite{d77}, which has the same value at $r=0$ but differs 
from the Kelbg potential at intermediate distances.
As was mentioned by Hansen \protect\cite{han75}, symmetry effects do not have a big
influence on the dynamical properties (although they give a major contribution
to the static properties, especially for the light mass components).
Using the Kelbg potential (\ref{kelbg}), we can immediately separate the short-range 
part of the interaction,
\begin{eqnarray}
U^{\rm sr}(r,T)=4\pi e^2 \!\left(\frac{-\exp(-r^2/\lambda^2)}
{r}+\frac{\sqrt\pi}{\lambda}{\rm erfc}(r/\lambda)\right),
\label{usr}
\end{eqnarray}
which has been calculated together with the first sum of Eq.~(\ref{ewald}) using 
the interpolation table.

The Kelbg potential contains just the lowest order quantum corrections (lowest 
order in $e^2$) and is, thus, accurate at small coupling, $\Gamma < 1$. 
Nevertheless, we expect that it correctly reproduces the influence of quantum effects 
also at intermediate coupling, $\Gamma \le 5$.
Further 
improvements are straightforward, e.g. by including exchange effects or by 
evaluating the full two-particle Slater sum.
We note that the described numerical procedure applies to such improved quantum 
pair potentials as well, even if they are not given analytically.

\subsection{Thermodynamic and dynamical quantities}
Solving Newton's equations with forces derived from the total potential 
$U^{\rm sr}+U^{\rm lr}$, we computed thermodynamic and static quantities, such as 
total energy and pair distribution function in usual manner. The results will be 
presented in the next section. Here we discuss some details on computation of the 
dynamical properties, as they require much more effort and computation time in order to 
achieve sufficient accuracy.

To obtain useful results for the dynamical structure factor, requires simulation results 
in a sufficiently broad range of wave numbers and frequencies. Natural units of the 
wave number and frequency are $1/{\bar r}$ and the plasma frequency 
$\omega_{pl}=\sqrt{4\pi e^2\rho/m}$, respectively, which will be used in the following. 
The minimum wave 
number $k_{min}$ depends on the size $L$ of the simulation box 
and thus, for a given density or coupling parameter, on the number of particles $N$.
One readily verifies that $k_{min} = 2 \pi / L = 2\pi (\rho/ N)^{1/3}$ or, 
using dimensionless wave numbers, $q_{min}=k_{min}{\bar r} = (6\pi^2/ N)^{1/3}$. Clearly, 
to reduce $k_{min}$ requires an essential increase of the number of particles in the 
simulation. 

The simulation accuracy can be further increased by taking advantage of the isotropy of 
the plasma in wave vector space. Indeed, in equilibrium, the density-density correlation 
function and dynamical structure factor should only depend on the absolute value of the 
wave vector. On the other hand, the simulations yield slightly different results 
for different directions of the wave vector. Averaging over all results
corresponding to the same absolute value of $\vec{k}$ allows to reduce the statistical 
error. For example, the minimum wave number $k_{min}$ corresponds to directions of 
$\vec{k}$ along either the x-, y- or z-axis, cf. Eq.~(\ref{rho_k}), so we can use the 
average of the three. The next larger value is $\sqrt{2}\,k_{min}$, 
corresponding to the diagonals in the x-y, x-z and y-z planes. The third value, 
$\sqrt{3}\,k_{min}$, corresponds to the space diagonal and is not degenerate;
 consequently it carries the largest statistical error. This is the main reason for 
the fluctuations of the 
numerical results for the wave vector dispersion, see for example Fig.~5.

Finally, to resolve the collective plasma oscillations, the duration of the simulations 
has to be much larger than the  plasma period. Also, increased simulation times 
leads to a reduction of the noise. We found that times of the order of 250  plasma 
periods are adequate.

\section{Numerical Results}\label{ch4}
We have performed a series of simulations for varying values of $\Gamma$ and $\rho\Lambda^3$, 
using the Coulomb and Kelbg potential. Also, time step and particle number have been 
varied until a satisfactory compromise between accuracy and simulation efficiency has 
been achieved. The parameters of the runs chosen for the figures below are summarized 
in Table~I. We mention that kinetic energy conservation in all runs (if velocity scaling 
was turned off) did not exceed $0.1 \%$. Also, the results for the total energy (not shown), in 
case of the Coulomb potential, agree very well with data from the literature. 

We first consider the pair distribution function $g(r)$ for varying
interaction  potentials and parameter values. Fig.~1 shows $g(r)$ for
three values of the coupling parameter, $\Gamma = 0.5, 1, 4$. As
expected, the Coulomb pair distribution function is close to the
Debye-H\"uckel limit  for small coupling, with increasing $\Gamma$, the
deviations, especially around $r={\bar r}$,  grow systematically. The
Kelbg pair distributions practically coincide with the  Coulomb
functions for $r>0.6\,{\bar r}$ but deviate from the latter at small
distances  of the order  of the thermal DeBroigle wave length $\Lambda$
where quantum effects are  important.  Clearly, with increasing
degeneracy, the ratio $\Lambda/{\bar r}$ increases, and the deviations 
extend to larger distances and grow in magnitude. With increasing
$\Gamma$, the deviations  become smaller since Coulomb effects dominate
the behavior at small distances. 

Let us now turn to the dynamical properties. In case of an OCP, charge and mass fluctuations 
are identical because of the rigid opposite charge background. In our simulations, 
we have calculated the density-density correlation function (\ref{dcf}) and, by 
numerical Fourier transformation, obtained  the dynamical structure 
factor $S(q,\omega)$ for several (from 6 to 10, depending on the simulation) wave 
numbers, the values of which are determined by the size of 
the simulation box L (see above). The value of the smallest wave number is given in Table~I.
The frequency dependence of $S(q,\omega)$ for several wave vectors is presented in Figs.~2-4 
for the Coulomb and Kelbg potentials. Also, the results of the mean-field models are shown.
The peak of the structure factor is related to the 
optical plasmon (Langmuir mode) of the electrons, its position shows the plasmon 
frequency $\Omega(k)$, its width - the damping of the mode. In the limit $k\rightarrow 0$, 
$\Omega(k)\rightarrow \omega_{pl}$ for all models. For increasing wave numbers, the 
width of the peak grows steadily, and it merges with the continuum of single-particle 
excitations, e.g. \cite{kker86,bonitz-book}, therefore, no results for larger wave 
numbers are shown. 
 
Consider now the results for the plasmon dispersion more in detail, cf. Fig.~5. 
First, we discuss the 
mean field results (\ref{s_mf}) which are calculated using the Vlasov and RPA polarizations, 
Eqs.~(\ref{p_vlasov}) and (\ref{p_rpa}), respectively.
The Vlasov result was computed using the formulas given in the review of Kugler \cite{kug73}, 
and for the RPA, a code was developed which accurately evaluates the pole integration in 
Eq.~(\ref{p_rpa}). Both approximations show the same general trend for small and 
intermediate wave numbers: with increasing wave number, 
the plasmon frequency and the damping increase. At large $q$, the dispersion 
exhibits a maximum and decreases again. In all situations, the RPA yields a 
slightly smaller frequency than the Vlasov result, whereas the damping values are 
very close to each other.

Let us now turn to the simulation results. The Coulomb and Kelbg simulations have been 
performed for exactly the same parameters, except for N and run time (cf. Table~I). (Notice that, in contrast 
to the Kelbg case, the Coulomb simulations depend only on $\Gamma$ which can be achieved 
by various combinations of density and temperature). Comparison of the two simulations shows, 
cf. Fig.~2, that 
the results for the structure  factors are very similar in case of small $\Gamma$. Peak positions and widths as well as the low and high frequency 
tails are very close to each other. The reason is obvious: since the potentials 
(and pair distributions, cf. Fig.~1) differ only at  a small interparticle distances 
of the order of $\Lambda$, differences  in the structure factor would show up only at 
$k > 2\pi/\Lambda$, 
which is about an order of magnitude larger than the wave numbers shown in Fig.~2.
There, the plasmon peak has already a width of the order of the frequency 
and no longer describes a well-defined collective excitation. 

It is now interesting to compare the simulation results to the theoretical approximations. 
The first observation is that the simulation peaks are significantly broader, cf. Fig.~2. This 
is obvious since the simulations fully include interparticle correlations missing 
in the mean-field results. Consequently, the plasmon damping contains collisional 
damping in addition to the Landau damping (which is the only damping mechanism in 
the mean-field models). Correspondingly, the plasmon peaks in the simulations are shifted to 
smaller frequencies. This effect grows with increasing wave number as well as 
with increasing coupling (see also Fig.~5).
We note that, in our simulations, this shift is observed for all wave numbers, which is 
in contrast to the result of Hansen [see Fig.~9 of Ref. \cite{han75}  for $q=0.6187$].
In other words, the plasmon dispersion curves from the MD simulations are lower than 
the mean-field result for all wave vectors $\vec{k}$, which is  seen more clearly in 
Fig.~5. As expected  the MD curves for the structure factor 
are much closer to the RPA than to the Vlasov result.

 In Fig.~5 we plot the optical plasmon dispersion curves for three values of the coupling 
parameter for the Vlasov and RPA dispersions together with the simulation results.
We further show the well-known analytical approximation to the Langmuir dispersion,
\begin{eqnarray}
\omega(q) = \omega_{pl}\left(1 + \frac{q^2}{\Gamma}\right)^{1/2}.
\label{vlasov_lim}
\end{eqnarray}
Clearly, this predicts a monotonically increasing dispersion. However, this approximation 
is valid only for $k<1/r_D$ and for $\Gamma < 1$. Let us now consider the simulation results 
which do not have this limitation. In Fig.~5 we show the MD results for a Coulomb potential 
and for the Kelbg potential for three values of the degeneracy parameter, 
$\rho\Lambda^3=0.1, 0.5, 1.0.$ 
One clearly sees that, for these parameters, the dispersion is positive, $d \omega(q)/dq > 0$,
up to wave numbers of the order of one over the mean interparticle distance. For larger 
$q$, the dispersion changes sign. This general trend is observed for 
the Coulomb and the Kelbg potential. On the other hand, with increasing quantum effects, 
$\rho\Lambda^3$, the 
deviations between the two potentials are growing, which becomes more pronounced as 
$\Gamma$ increases, cf. the curves for $\Gamma=1$ and $\Gamma=4$:  
the dispersion in case of the Kelbg potential shows a softer increase
with increasing wave number and reaches a lower maximum approximately at the 
same wave number as in the Coulomb case. We mention that this sign change of the dispersion 
has not been reported by Hansen \cite{han73}.
Comparing the simulations with the mean-field results, we again see 
that the MD dispersions proceed lower than the mean field results, and this effect 
grows with increasing $\Gamma$ and increasing wave number. Once more, we confirm that 
the RPA dispersion is much closer to the MD result than the Vlasov dispersion, at least 
for $\Gamma \le 0.5$. [As mentioned above, the simulation results for the dispersion 
show certain statistical fluctuations due to the varying accuracy of the results for the 
different wave numbers]. 

Let us now consider the plasmon damping more in detail. Fig.~6, shows the 
damping (full width at half maximum of the plasmon peak of the structure factor) as a 
function of wave number. It is interesting to compare with the familiar analytical 
expression from the Vlasov theory, e.g. \cite{abr-book},
\begin{eqnarray}  
\delta (\kappa) =  \sqrt{\frac{\pi}{8}} \,\frac{\sqrt{1+3\kappa^2}}{\kappa^3} \,
 e^{-\frac{1}{2 \kappa^2}- \frac{3}{2}
}  
\label{vlasov_damp}
\end{eqnarray}
where $\kappa \equiv k r_D$ is the dimensionless wave number in units of the inverse 
the Debye radius $r_D$ given in Table~I. Formula (\ref{vlasov_damp}) is derived under the condition 
that the damping is much smaller then the frequency [$\delta(q) \ll \omega(q)$], and is 
limited to small wave numbers $\kappa \ll 1$. 
As expected, the damping given by formula (\ref{vlasov_damp}) which is only
Landau damping, is much smaller than the damping found in the simulations, as the latter 
contain the full collisional damping also. Obviously, for small coupling and 
small $q$,  Eq. (\ref{vlasov_damp}) shows the correct trend. However, deviations 
increase rapidly with growing coupling parameter. Furthermore, the simulations which are 
not limited to small wave numbers, show a qualitatively different behavior at large $q$:
a monotonic increase of the damping. Interesingly, with increasing $\Gamma$ the
 damping is reduced, cf. Figs. 6a,b.

Finally, we try to extend the analytical result for the plasmon dispersion, Eq. (\ref{vlasov_lim}), 
to larger $\Gamma$ and to include quantum effects. To this end, we used the MD data with the 
Kelbg potential to construct an improved fit of the form
$\omega(q) = \omega_{pl}\left(1 + a q^2 + b q^4\right)^{1/2}$. The result is shown in 
Fig.~7 for $\Gamma=1$ and $\Gamma=4$. 
Due to the large fluctuations in the dispersion data and the increasing damping for large 
wave numbers, we used a weighted fit where the smallest $q-$values had the largest weight 
and the statistical errors of the individual points have been taken into account.
Table~II contains the resulting fit parameters. 
According to this data both parameters $a$ and $b$ depend on
$\Gamma$ and $\rho\Lambda^3$. The parameter $a$ is close to $1/\Gamma$ in agreement 
with Eq. (\ref{vlasov_lim}), but with increasing $\Gamma$, deviations are growing, compare 
Table~II. We see no systematic influence of quantum effects on the parameter $a$ for $\Gamma=1$.
Noticeable effects show up for $\Gamma=4$, where increased degeneracy leads to a reduction 
of the coefficient $a$. The second fit parameter allows to qualitatively reproduce the change 
of the sign of the dispersion. The overall agreement is satisfactory for wave numbers up to 
the inverse mean interparticle distance up to which the plasmons are comparatively 
weakly damped.

\section{Discussion}\label{ch5}
We have presented classical molecular dynamics simulations of the dielectric 
properties of a one-component plasma at intermediate coupling and degeneracy, 
$\Gamma\le 4$ and $\rho\Lambda^3\le 1$. While classical MD simulations can be extended to 
very large values 
of $\Gamma$, they have limited applicablility to quantum plasmas. We used, as an 
effective quantum pair potential, the Kelbg potential which correctly describes quantum 
diffraction effects for small $\Gamma$. 
In general, we found that the simulation results for the dielectric properties and the 
plasmon dispersion with the 
Coulomb and the Kelbg potential are rather close, but start to deviate from each other 
as $\Gamma$ increases. Nevertheless, the use of the Kelbg potential is preferable. 
It correctly treats close collisions, i.e. the two-particle interaction at distances smaller than 
the DeBroglie wavelength. This is of even higher importance in the case of two-component 
plasmas where the Kelbg potential allows to avoid the collapse of oppositely charged 
particles. Therefore, the present investigation should be important for future work on 
two-component plasmas.
Finally, we mention that the Kelbg potential is only the first term of a 
$\Gamma$ expansion. Therefore, for $\Gamma > 1$ the account of 
higher order corrections to the quantum diffraction effects is necessary. Work on this 
subject is in progress.

\begin{table}
\caption{Parameters of the molecular dynamics simulations with the Kelbg 
potential. Numbers in 
parentheses refer the runs with Coulomb potential.
\\
}
\begin{tabular}{|c|c|c|c|c|c|c|c|c|}
$\Gamma$ & $\rho\Lambda^3$ & $\rho$, [cm$^{-3}$] & T, [K] & $\omega_{pl}$,
 [fs]$^{-1}$ & $ r_D/ \bar{r}$ & 
$N$ & $k_{min}{\bar r}$ & run time, [$T_{pl}$]\\
\hline 
0.5 & 0.1 & 9.12 $\cdot 10^{21}$ & 1.126 $\cdot 10^5$ & 5.387 & 0.816 &
500(250)& 0.491(0.619) & 515(341) \\ 

0.5 & 0.5 & 2.28$\cdot 10^{23}$ & 3.292$\cdot 10^5$ & 26.940 & & 
400(250) & 0.529(0.619) & 429(429) \\ \hline

 1.0 & 0.1 & 1.14$\cdot 10^{21}$ & 2.228$\cdot 10^4$ & 1.905  & 0.577 & 
250 & 0.619 & 290(327)\\

1.0 & 0.5 &2.85$\cdot 10^{22}$ & 8.23 $\cdot 10^4$ & 9.524 & &
250 & 0.619 & 682(682)\\

1.0 & 1.0 &1.14$\cdot 10^{23}$ & 1.31 $\cdot 10^5$ & 19.048 & &
250 & 0.619 & 477\\ \hline

 4.0 & 0.1 & 1.78$\cdot 10^{19}$ & 1.76$\cdot 10^3$ & 0.238  & 0.289 & 
250 & 0.619 & 570(227)\\

4.0 & 1.0 &1.78$\cdot 10^{21}$ & 8.17 $\cdot 10^3$ & 2.381 & &
250 & 0.619 & 716\\
\end{tabular}
\end{table}

\begin{table}
\caption{Fit parameters of the Langmuir dispersion curves shown on Fig.~\ref{fig7}. 
The fit equation was taken in the form of $ \omega(q)/\omega_{pl} = (1 + a q^2 + b q^4)^{1/2}$.   
Parameters of the fit for $\Gamma$ = 1 and $\rho \Lambda^3$ = 0.1 are less reliable, 
because of the absence of data for big wave vectors, cf. Table I.\\
}
\begin{tabular}{|c|c|c|c|}
$\Gamma$ & $\rho\Lambda^3$ & a & b \\
\hline 
1.0&0.1& 1.013 $\pm$ 0.031  & -0.260 $\pm$ 0.023\\ 
1.0&0.5& 1.074 $\pm$ 0.041  & -0.288 $\pm$ 0.013\\
1.0&1.0& 0.975 $\pm$ 0.055  & -0.259 $\pm$ 0.018\\
\hline
4.0&0.1& 0.169 $\pm$ 0.015  & -0.034 $\pm$ 0.006\\
4.0&1.0& 0.121 $\pm$ 0.007  & -0.025 $\pm$ 0.003\\
\end{tabular}
\end{table}

\begin{figure}[htb]
\begin{picture}(300,170)
\put(0,-300){\psfig{file=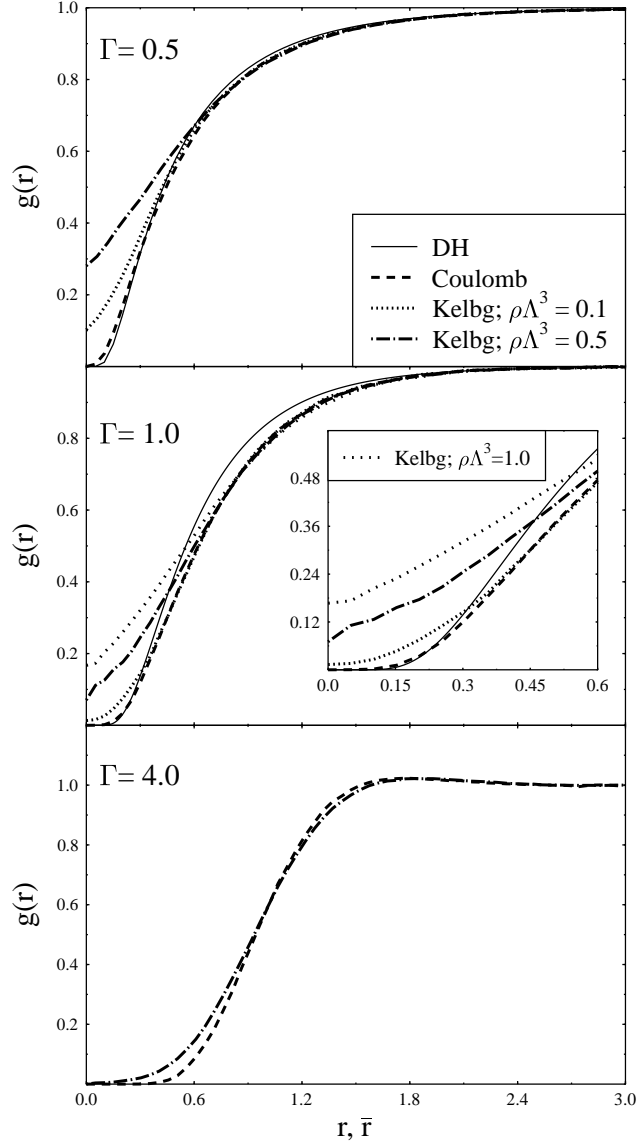,width=120mm}}
\end{picture}
\put(-300,-450){\caption {Pair distribution functions for 
$\Gamma $= 0.5 (upper figure), $\Gamma $= 1.0 (middle figure),
4.0 (lower figure), and $\rho \Lambda^3$ = 0.1, 0.5, 1.0 for systems with Coulomb and
Kelbg potential. Further, the Debye-H\"uckel (DH) limit is shown (solid line).
Line styles are the same in all three figures.
Inset in the middle Fig. shows the influence of the degeneracy at small distances. 
The result for $\Gamma$ = 4.0 , $\rho \Lambda^3$ = 0.1
with Kelbg potential are not distinguishable from the  Coulomb result and are not plotted.}
\label{fig1} }
\end{figure}

\begin{figure}[htb]
\begin{picture}(300,140)
\put(0,-300){\psfig{file=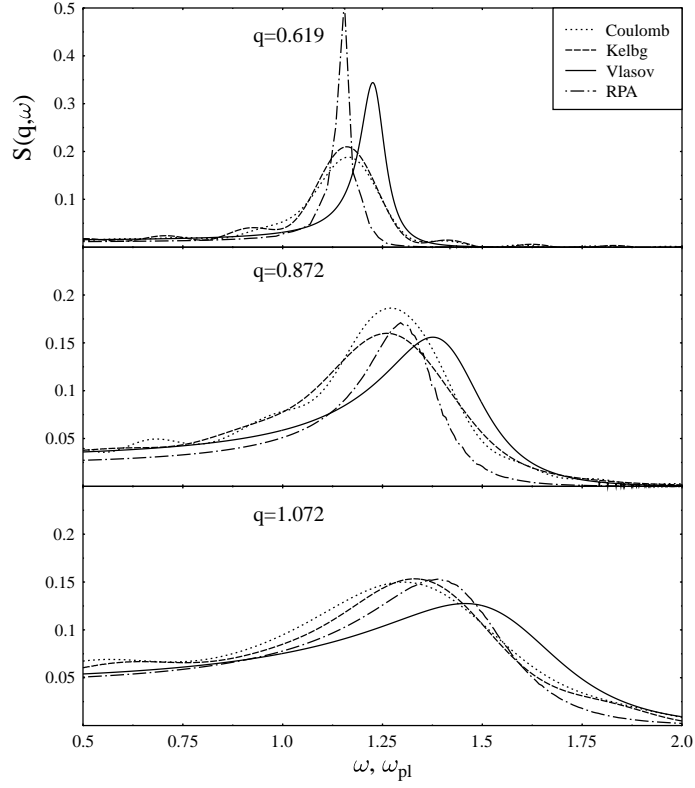,width=120mm}}
\end{picture}
\put(-300,-250){\caption{Dynamical structure factor for an OCP at $\Gamma$ = 1 and $\rho
\Lambda^3$ = 0.1 from MD simulations with Coulomb and Kelbg potentials. 
In addition, Vlasov and RPA results are shown. The wave numbers are shown in the figures 
in units of $\bar{r}$, i.e. $q = k\bar{r}$. }
\label{fig2} 
}
\end{figure}

\begin{figure}[htb]
\begin{picture}(300,140)
\put(0,-300){\psfig{file=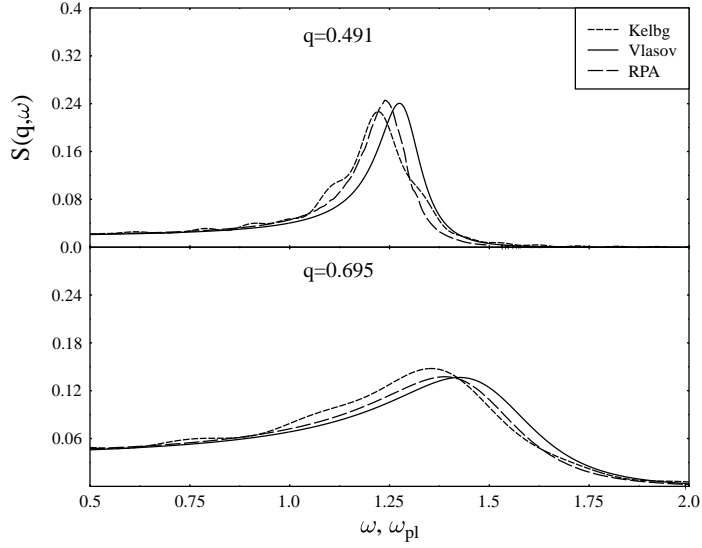,width=120mm}}
\end{picture}
\put(-300,-120){\caption{Same as Fig.~2, but for $\Gamma$ = 0.5 and
$ \rho \Lambda^3$ = 0.1. The values of the wave numbers differ from Fig.~1 due to 
the different particle numbers, cf. Table~I.} 
\label{fig3}
}
\end{figure}

\begin{figure}[htb]
\begin{picture}(300,140)
\put(0,-300){\psfig{file=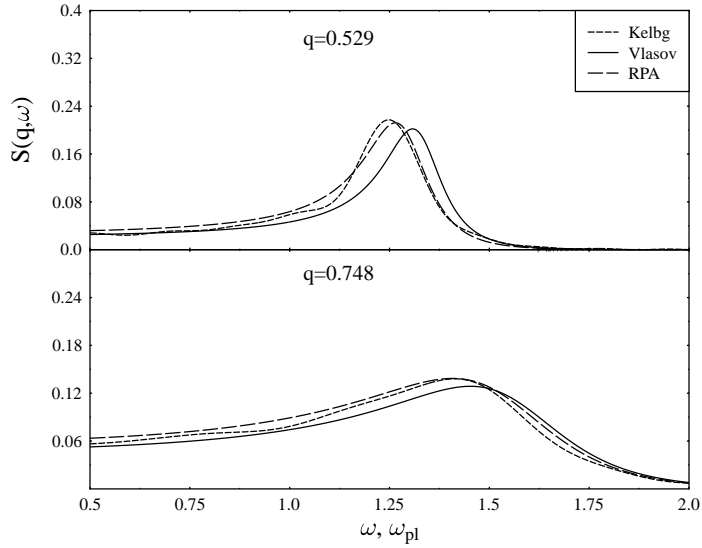,width=120mm}}
\end{picture}
\put(-400,-110){\caption{Same as Fig.~3, but for $\rho\Lambda^3$ = 0.5.}
\label{fig4} }
\end{figure}

\begin{figure}[htb]
\begin{picture}(400,140)
\put(0,-400){\psfig{file=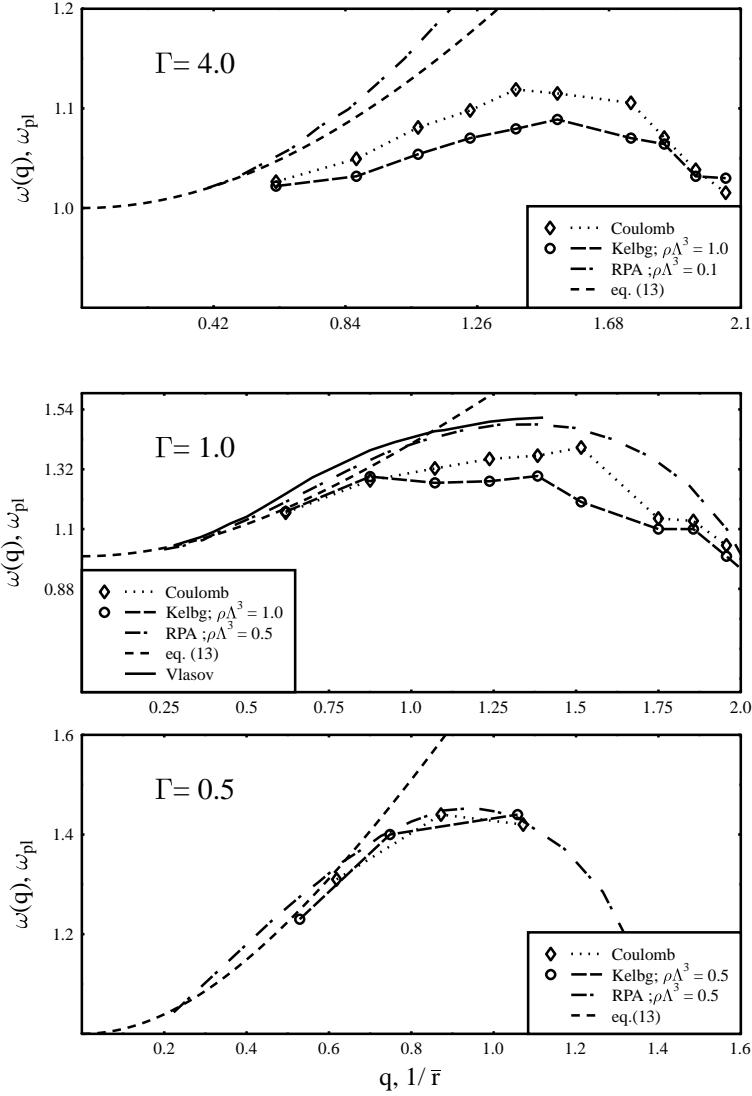,width=120mm}}
\end{picture}
\put(-400,-480){\caption{Optical plasmon dispersion for various coupling and degeneracy 
parameters from MD simulations with Coulomb and quantum potentials. Also shown are results of 
the Vlasov and RPA approximations, and of the analytical approximation of  Eq.~(\ref{vlasov_lim}). 
For $\Gamma=$ 4.0 and $\rho \Lambda^3$ = 0.1 (upper graph) the MD simulations with Kelbg potential 
and the RPA curve are not shown since they almost coincide with the Coulomb simulation and 
the Vlasov curve, respectively. }\label{fig5} }
\end{figure}

\begin{figure}[htb]
\begin{picture}(400,140)
\put(0,-400){\psfig{file=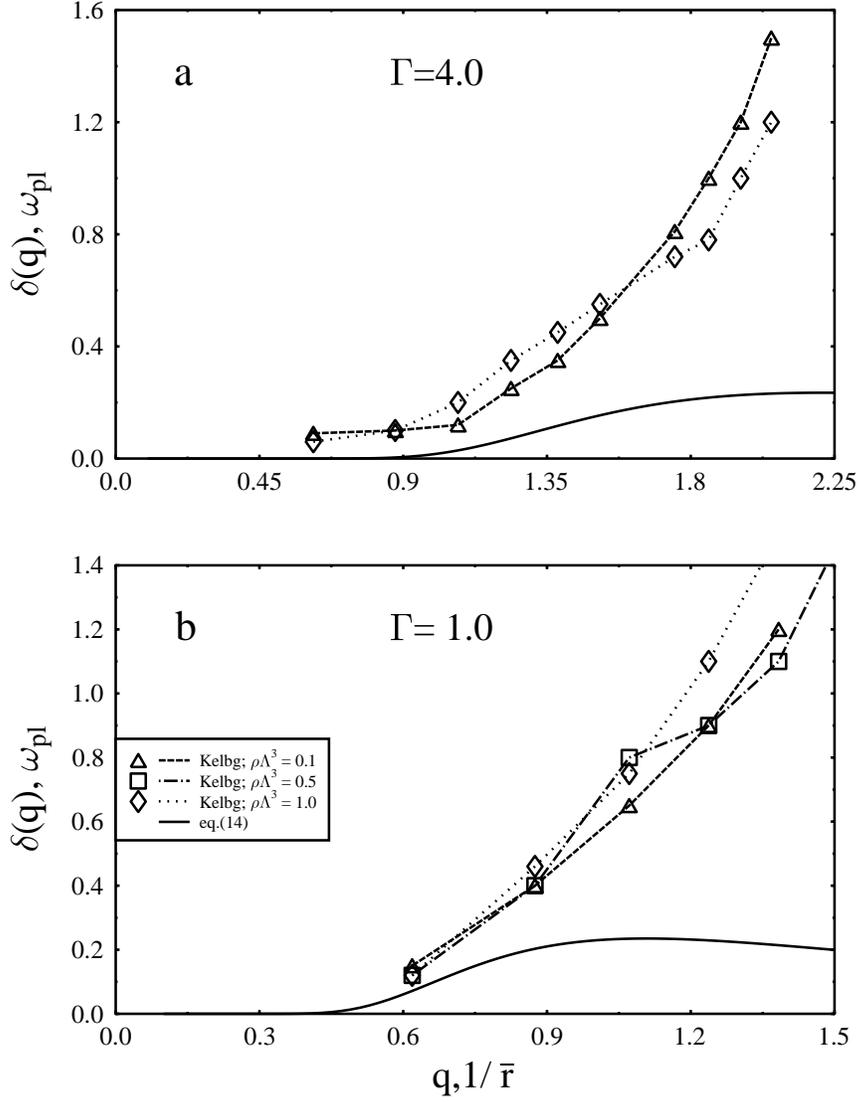,width=120mm}}
\end{picture}
\put(-400,-450){\caption {Damping of Langmuir waves from MD simulations with the Kelbg 
potential for various values of $\Gamma$ and $\rho \Lambda^3$. Solid lines are the analytical 
small damping limit of the Vlasov theory, Eq.~(14).} \label{fig6}}
\end{figure}

\begin{figure}[htb]
\begin{picture}(400,140)
\put(0,-400){\psfig{file=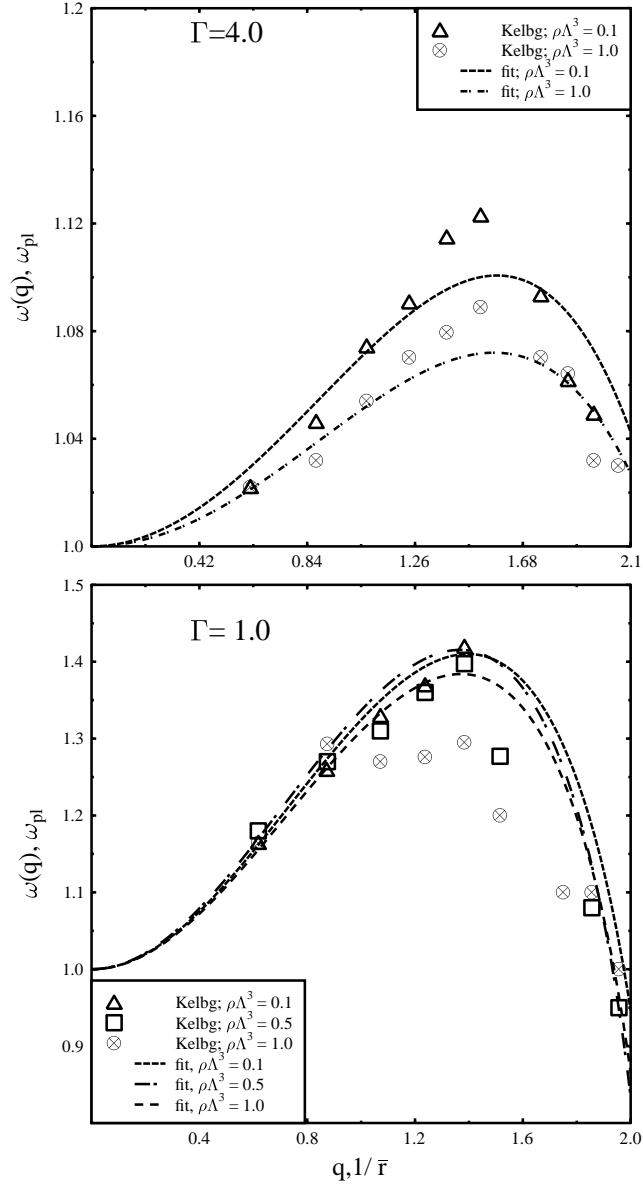,width=120mm}}
\end{picture}
\put(-400,-450){\caption{Dispersion of Langmuir oscillations from MD simulations with the 
Kelbg potential for various values of the coupling and degeneracy. 
Symbols are MD results, lines the best fits to the low wave number part 
($q < 1/{\bar r}$), the fit formula and parameters are given in Table~II. 
}\label{fig7} }
\end{figure}

\end{document}